# Morphological Profiling for Drug Discovery in the Era of Deep Learning

Qiaosi Tang[1], Ranjala Ratnayake[2], Gustavo Seabra[2], Zhe Jiang[3], Ruogu Fang[3,4], Lina Cui[2], Yousong Ding[2], Tamer Kahveci[3], Jiang Bian[5], Chenglong Li[2], Hendrik Luesch[2], Yanjun Li[2]*

[1] Calico Life Sciences, South San Francisco, CA 94080, USA
[2] Department of Medicinal Chemistry, Center for Natural Products, Drug Discovery and Development, University of Florida, Gainesville, FL 32610, USA
[3] Department of Computer & Information Science & Engineering, University of Florida, Gainesville, FL 32611, USA
[4] J. Crayton Pruitt Family Department of Biomedical Engineering, Herbert Wertheim College of Engineering, University of Florida, Gainesville, FL 32611, USA
[5] Department of Health Outcomes and Biomedical Informatics, College of Medicine, University of Florida, Gainesville, FL 32610, USA

* Correspondence: Yanjun Li
Email address: yanjun.li@ufl.edu

## Abstract

Morphological profiling is a valuable tool in phenotypic drug discovery. The advent of high-throughput automated imaging has enabled the capturing of a wide range of morphological features of cells or organisms in response to perturbations at the single-cell resolution. Concurrently, significant advances in machine learning and deep learning, especially in computer vision, have led to substantial improvements in analyzing large-scale high-content images at high-throughput. These efforts have facilitated understanding of compound mechanism-of-action (MOA), drug repurposing, characterization of cell morphodynamics under perturbation, and ultimately contributing to the development of novel therapeutics. In this review, we provide a comprehensive overview of the recent advances in the field of morphological profiling. We summarize the image profiling analysis workflow, survey a broad spectrum of analysis strategies encompassing feature engineering- and deep learning-based approaches, and introduce publicly available benchmark datasets. We place a particular emphasis on the application of deep learning in this pipeline, covering cell segmentation, image representation learning, and multimodal learning. Additionally, we illuminate the application of morphological profiling in phenotypic drug discovery and highlight potential challenges and opportunities in this field.

# 1 Introduction

Phenotypic drug discovery (PDD) plays a crucial role in the field of drug discovery. In contrast to target-based drug discovery (TDD), where compounds are designed to interact with known target molecules, PDD takes a target-agnostic approach and focuses on phenotypic effects of compound treatment in disease-relevant biological systems[1,2] (Figure 1A, 1B). This strategy uses reference compounds with treatment class annotations to uncover previously unknown mechanism of actions (MOAs) of the test compounds. To date, PDD has made a significant contribution to the development of first-in-class drugs and the discovery of novel therapeutic opportunities[1,2]. For example, PDD is the primary approach in natural products discovery and the basis for identification of new targets and/or MOAs. Natural products are all bioactive, and the most effective way to multiplex and assign function is through phenotypic screening, particularly by analyzing related biased and unbiased nuances from high-content imaging[3-6].

Automated microscopy and image analysis have enabled high-throughput image-based assays for PDD[1,7]. The two approaches, namely high-content screening (HCS) and morphological profiling, are both based on large-scale imaging experiments, yet they are distinct in strategy (Figure 1C, 1D). In HCS, feature measurements are limited to specific phenotypes related to perturbations. In other words, HCS adopts a pre-defined strategy to capture a limited number of biological properties of interest, effectively examining expected phenotypic effects. In contrast, morphological profiling (also known as image-based profiling or cytological profiling) takes an unbiased approach to capture high-dimensional image data, consisting of hundreds to thousands of cellular features. Conventionally, bioimage informatics tools measure these features, which span a range of morphological properties such as size, shape, texture, and intensity. The extracted features are then used to generate phenotypic signatures, requiring computational models for clustering and predicting perturbation bioactivity similarity[7,8]. Thus, this approach not only provides a comprehensive morphological profile in an unbiased manner, but also allows for detecting subtle or novel phenotypes for pharmaceutical discovery.

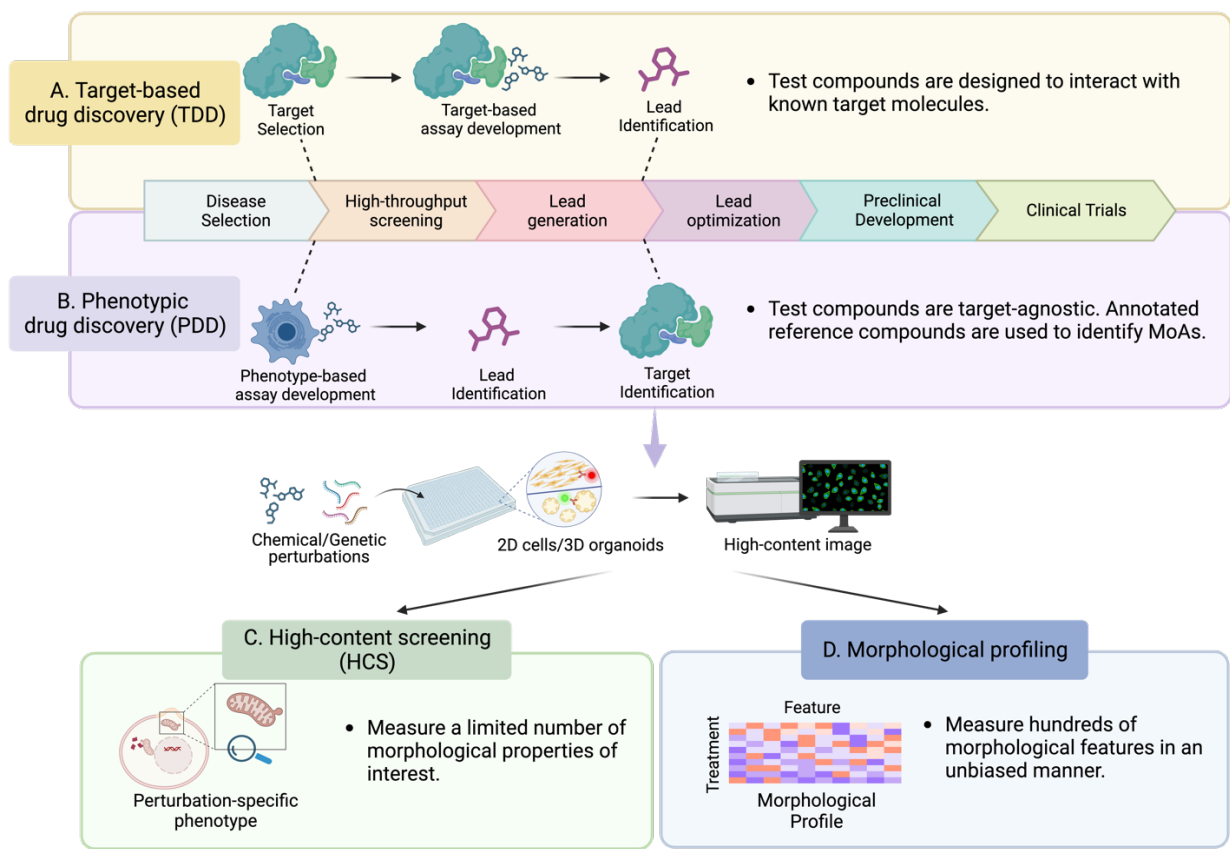

**Figure 1. Early-stage drug discovery approaches.**
Figures A and B illustrate two primary approaches in drug discovery: target-based drug discovery (TDD) and phenotypic drug discovery (PDD). **A)** TDD starts with a known drug target, and a target-based assay is established to evaluate the effect of compound-target interaction. **B)** In contrast, PDD employs a target-agnostic strategy, screening compounds to determine whether a phenotype-of-interest is induced. Due to its unbiased nature, a target identification step is required. Within the context of PDD, **C)** high-content screening (HCS) and **D)** morphological profiling are two commonly used approaches. The major difference is **C)** uses a limited number of perturbation-specific phenotypes as assay readout, whereas **D)** morphological profiling obtains cellular feature representation with an unbiased approach.

Among the vast array of fields where Artificial Intelligence (AI) can be applied, medicine stands out as a domain where significant potential and substantial challenges coexist[9]. AI has revolutionized various aspects of healthcare and biomedical research, including precision medicine[10], medical image analysis[11], and drug discovery[12,13]. As a dominant technique in AI, deep learning is a part of broader machine learning area that uses deep neural networks to learn representations from raw data in a data-driven manner, often without the need for feature engineering[14]. In the context of drug discovery, deep learning has emerged as a potent tool, enabling efficient development of novel therapeutics through various applications, such as target identification[15,16], protein structure prediction[17,18], drug-target interaction prediction[19-21], *de novo* drug design[22,23], molecular property prediction[24,25] and biological image analysis for PDD[26,27]. In recent years, computer vision has made significant breakthroughs in various subdomains because of the advancements in deep learning, such as object detection, sematic segmentation, transfer learning, image generation, and more, leading to a profound transformation of image-based profiling analysis in efficiency and performance, thereby expediting drug discovery and reducing computational cost[27,28].

In this review, we aim to provide a comprehensive overview of the extant computational approaches employed in morphological profiling with a particular emphasis on the deep learning applications. We primarily focus on the analytical pipeline of Cell Painting high-content image data given its wide application in academic research and pharmaceutical industry. We start with an introduction of Cell Painting image analysis workflow with conventional feature-engineering approach (also known as 'handcrafted' representation). For the major focus, we provide a thorough summary of recently proposed deep learning approaches in advancing this analytical pipeline, including microscopic image cell segmentation, representation learning from high-content fluorescent images and multimodal learning to integrate chemical structure and omics data for MOA prediction. With concrete examples in cutting-edge applications of constructing gene function network, morphodynamics profiling, *de novo* hit design, compound MOA prediction in organoids and natural product annotation, we conclude with our perspectives on future directions in advancing morphological profiling with deep learning solutions.

## 2 Deep learning in morphological profiling analytical pipeline

### 2.1 Cell Painting and benchmark datasets

A state-of-the-art assay for morphological profiling is known as Cell Painting[29]. The canonical protocol on adherent monolayer cells uses six fluorescent dyes to characterize eight cellular components or organelles, and images the fixed and stained cells in five channels on a high-throughput microscope[30]. Recent optimization efforts have further improved the assay's capability in phenotype detection[31]. Whereas canonical Cell Painting captures cellular morphology in snapshots, technical advances now enable live-cell imaging, such as using reporter cell lines that carry organelle or pathway marker with fluorescent tag. This allows for capturing morphological profiles in dynamics[32].

Over the past decade, morphological profiling efforts from academia and pharmaceutical industry have produced several publicly available Cell Painting datasets. These include 1) the Broad Bioimage Benchmark Collection (BBBC) with compound and genetic perturbations [33-35], 2) The Image Data Resource (IDR) with both HCS images and time-lapse images[36], and 3) the RxRx datasets released from Recursion with compounds, genetic and viral transduction perturbations, and 4) the CytoImageNet dataset curated from 40 openly available and weakly-labeled microscopy images[37]. Notably, the Joint Undertaking in Morphological Profiling Cell Painting (JUMP-CP) Consortium has recently been established as the largest public reference Cell Painting dataset[38], including images from more than 116,000 chemical perturbations and over 15,000 genetic perturbations on human osteosarcoma cells (U2OS), which were systematically acquired from 12 data-generating centers[38]. An extension to this dataset, labeled CPJUMP1, has been curated to include pairs of chemical and genetic perturbations that both target the same genes in the settings of U2OS and human lung carcinoma epithelial cells (A549)[39]. These public reference datasets have been broadly used to train machine learning and deep learning models for compound bioactivity prediction, and image representation learning for feature embedding. Details of these datasets are summarized in Table 1.

**Table 1. Publicly available cellular microscopic image datasets for model training and evaluation.**

| Data set | Description | URL | Reference |
|---|---|---|---|
| The Broad Bioimage Benchmark Collection (BBBC) | A collection of image datasets from image-based profiling and other assays annotated with different types of ground truth. | https://bbbc.broadinstitute.org/image_sets | Ljosa 2013[33] |
| Recursion datasets (RxRx) | Image datasets with different perturbation modalities such as genetic, small molecule and viral infection perturbations. | https://www.rxrx.ai/ | Sypetkowski 2023[40] |
| Image Data Resource (IDR) | A public repository of datasets from image-based assays. | https://idr.openmicroscopy.org/cell/ | Williams 2017[36] |
| JUMP Cell Painting datasets (JUMP-CP) | A multi-center image dataset of U2OS cells under genetic and compound perturbations. | https://registry.opendata.aws/cellpainting-gallery/ | Chandrasekaran 2023[38] |
| CPJUMP1 | An image dataset of matched chemical and genetic perturbations targeting the same genes in U2OS and A549 cells. | https://broad.io/neurips-cpjump1 | Chandrasekaran 2022[39] |
| CytoImageNet | A dataset curated from the above publicly available microscopic images with weak labels for bioimage transfer learning. | https://www.kaggle.com/datasets/stanleyhua/cytoimagenet | Hua 2021[37] |

**2.2 An overview of image-based profiling data analysis**

An accurate, efficient, and generalizable imaging data analysis workflow is critical for morphological profiling. Established methods and best practices to produce high-quality image-based profiles have been comprehensively documented[41-44]. However, the past few years have witnessed significant strides in the application of deep learning approaches, achieving unprecedented performance levels in various computer vision projects. These advances have extensively enriched the field of morphological profiling (Figure 2). In this section, we present an overview of the critical stages in morphological profiling data analysis, with particular emphasis on recent advances (Figure 3).

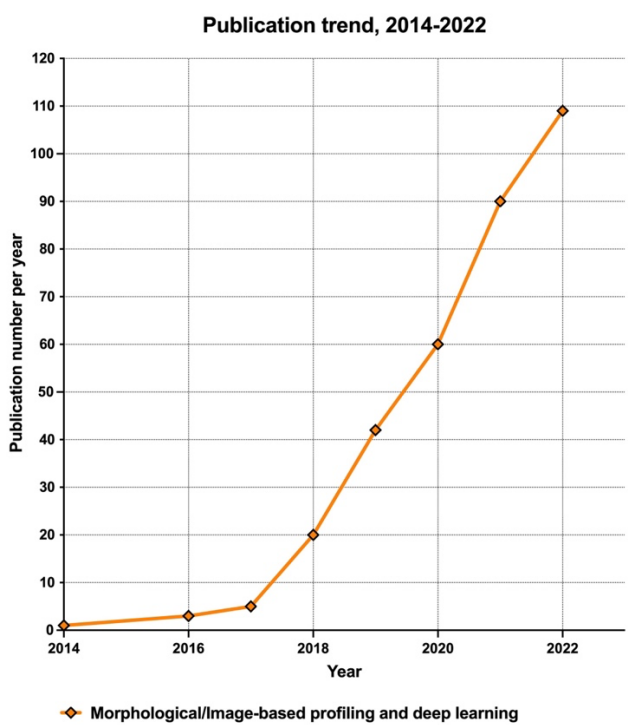

**Figure 2. Recent publication trend of morphological profiling with deep learning.**
Pubmed trend demonstrates a growing number of indexed publications on morphological profiling with deep learning, including the keywords 'Deep learning' with 'Morphological profiling' or 'Image-based profiling'. This trend is analyzed from 2014 to 2022.

*Stage 1: Feature representation*

Measuring variations in cell morphology upon perturbation relies on generating effective representations for cellular images. Conventionally, this task is implemented by feature engineering approaches. Bioimaging software like CellProfiler is commonly used to extract predefined features such as cell shape, size, and texture from fluorescent microscopy images[45].

While this approach provides biologically insightful results, it requires image preprocessing and manual adjustment of parameters for every new experiment setup[41,43]. Also, single cell segmentation is typically required, which will be described in detail in Section 2.3. Recently, instead of representing cell morphology with raw reads from CellProfiler feature extraction, a new metric, the equivalence score (Eq. score), has been proposed to represent the similarities of different treatments relative to baseline non-treatment controls. The Eq. score is computed by Partial Least Squares (PLS/OPLS) regression from morphological feature raw reads. The use of Eq. scores has been demonstrated to reduce noise and enhance biological signal recovery in treatment groups compared to non-treatment controls[46].

Alternatively, deep neural networks such as pre-trained convolutional neural networks (CNNs) can learn representation directly from a full-field microscopy image without the need for single-cell segmentation[47,48]. Further, generative adversarial network-based (GAN) models and variational autoencoder (VAE) framework have been proposed to improve the interpretation of cellular structural variations that drive morphological differences[49-51] and to predict morphological responses to perturbations[52]. In scenarios where ground truth labels are noisy or missing, strategies such as weakly supervised learning (WSL) have been utilized to learn feature embeddings at both single-cell and full-field resolutions[53,54]. These advances and benefits from deep learning-based analysis approaches will be further discussed in section 2.4.

### *Stage 2: Morphological profile generation*

Once features are extracted from single cells or field images, these measurements will be aggregated into a single feature vector for well-level (also known as treatment-level or population-level) representation. The morphological profile generated from this stage will enable downstream well-level analysis. Aggregation via the median of each feature across all cells has been a prevalent approach[41]. However, such simple aggregation like the median approach does not account for cellular heterogeneity, wherein different phenotypes can be discerned from different cell subpopulations, even under the same treatment[55]. Considering this, subpopulation identification by clustering cells into different phenotypic subgroups based on their single-cell profiles can improve the understanding of phenotypic transition dynamics[41]. Notably, this aggregation approach of defining subpopulation has also been applied in organoid morphological profiling[56].

*Stage 3: Classification to predict compound bioactivity*

With the aggregated treatment-level morphological profiles, a common machine learning task is to predict MOA or toxicity of query perturbagens based on the known morphological profiles of the reference library[42]. This is most commonly achieved by building a feature-based machine learning model such as nearest neighbor classifier, random forests and Bayesian matrix factorization[41,42,57] on top of the extracted morphological profiles, which are typically represented as feature vectors. Precision, recall and accuracy measurements obtained via holdout test set are used as evaluation metrics for the classification performance[41]. With these supervised machine learning algorithms, query perturbagens can be classified into pre-defined, annotated classes[42].

*Stage 4: Clustering to infer treatment association*

Aggregated morphological profiles can be used to infer treatment-level associations. This task is accomplished by employing hierarchical unsupervised clustering algorithms, predicted on the similarity of morphological profiles[42]. A phenotypic similarity matrix of all pair-wise similarities between morphological profiles is computed for similarity-based clustering[42]. The statistical significance of the results can be evaluated by bootstrapping[41].

Notably, deep learning techniques also facilitate an end-to-end learning schema, integrating all the aforementioned stages into a singular, unified process. Within this framework, deep learning models can directly undertake the phenotypic classification and clustering tasks using raw high-content images, circumventing the explicitly image feature representation, morphological profile generation, and other intermediate steps. This end-to-end learning schema will be elaborated in Section 2.4.

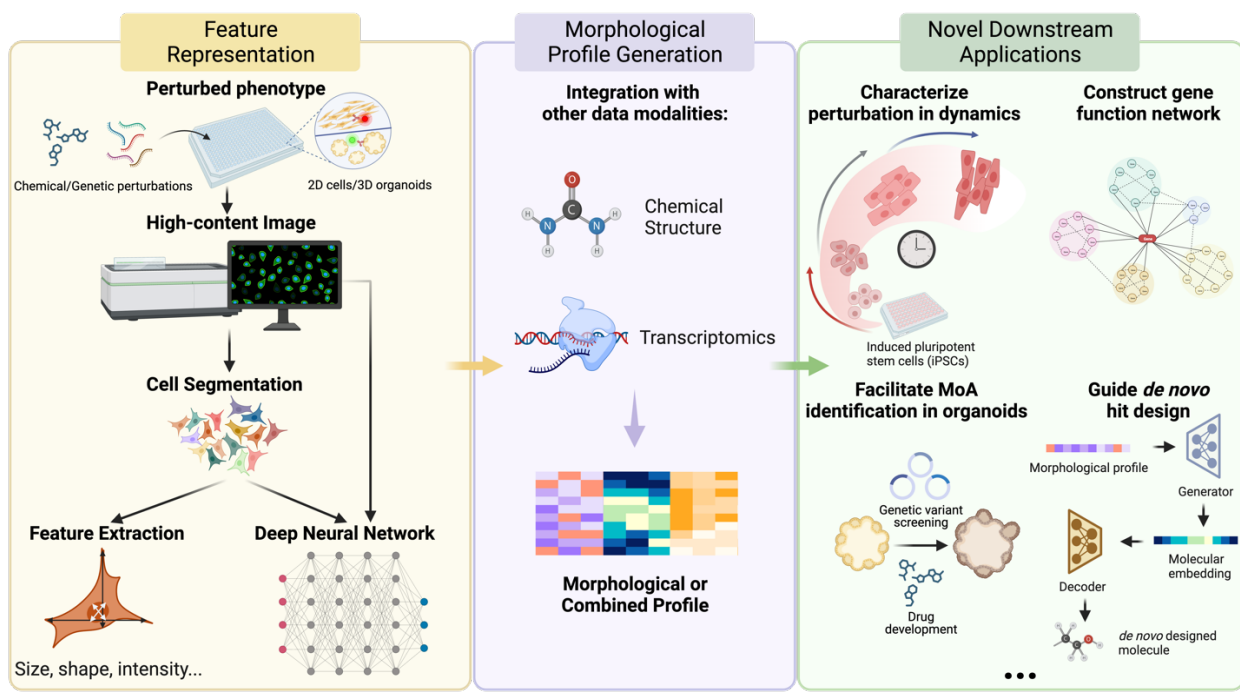

**Figure 3. Schematic workflow of morphological profiling.**
After cells are perturbed and stained, fluorescent images are taken to capture cellular morphology. Single cells are detected and segmented. At the single-cell level, morphological features can be achieved with image analysis software to extract pre-defined features. Alternatively, feature vectors can be obtained through representation learning with a deep neural network. Features from single cells are subsequently aggregated into a treatment-level morphological profile. Certain deep learning models allow end-to-end learning, eliminating the need for cell segmentation. The resulting morphological profile is then applied to downstream tasks such as classification for MOA prediction and clustering for treatment association inference **(Left Panel)**. Additionally, other profile modalities, such as chemical structure and transcriptomic profile, can be integrated with the morphological profile to enhance downstream analysis **(Middle Panel)**. Altogether, these efforts enable many novel downstream applications, such as characterizing perturbation impacts in dynamics, constructing gene function network to map genotype-phenotype relationship, identifying compound MOAs in 3D organoid model, and guiding *de novo* hit design **(Right Panel)**.

## 2.3 Deep learning-facilitated feature engineering methods for image analysis

Although many image analysis software programs were developed prior to the emerging of deep learning, progress in the field of computer vision, a subdomain of deep learning, has significantly enhanced image analysis tools. In this section, we will start by providing a concise overview of the advances in open-source image analysis software and then focus on how deep learning enables cell segmentation, a crucial step of image analysis.

### 2.3.1 Open-source scientific image software and resource advances

Open-source image analysis software and packages have long been benefiting the bioimage analysis community. Two leading platforms, namely ImageJ and CellProfiler, offer interactive image processing functionality for single image processing[58]. ImageJ, originally named National Institute of Health (NIH) Image, was developed in 1987. Over the past decades, professional developers and bench biologists have contributed and benefited from its powerful plugins and macros[59]. The core ImageJ team has significantly enhanced its interoperability with external applications[60]. Although ImageJ is designed for single-image processing, it is possible to integrate with KoNstanz Information MinEr (KNIME) for large-scale analysis[58,61]. CellProfiler, on the other hand, was developed to meet the needs of large-scale image analysis pipelines[45]. The CellProfiler ecosystem has seen continued enhancements, with the most recent release of CellProfiler 4.0 for improved performance in speed and utility[62]. To further implement cutting-edge methods with deep learning models and to better address the diverse needs of bioimage analysis, the Center for Open Bioimage Analysis (COBA) has been established to interoperate the strengths of the ImageJ and CellProfiler platforms[58]. Example protocols to integrate both platforms into a single analysis pipeline have been described[58]. In addition to ImageJ and CellProfiler, we have also summarized other open-source image analysis software and tools in Table 2.

**Table 2. A selection of open-sourced image analysis tools.**

| Tools | Website | Function |
| --- | --- | --- |
| AGAVE | https://www.allencell.org/pathtrace-rendering.html | 3D volume image viewer. |
| AICSImageIO | https://github.com/AllenCellModeling/aicsimageio | Python module for image reading, writing and metadata conversion. |
| Aydin | https://github.com/royerlab/aydin | Python module for Image denoising. |
| Bio-Formats | https://www.openmicroscopy.org/bio-formats/ | Software for reading and writing image data and metadata. |
| BioImageIO | https://bioimage.io/#/ | Deep learning model repository for image segmentation |
| Cellpose | https://www.cellpose.org/ | Deep learning model for image segmentation. |
| CellProfiler | https://cellprofiler.org/ | Software for automated feature extraction on large-scale image dataset. |
| CLIJ | https://clij.github.io/ | GPU-accelerated image processing library for Fiji/ImageJ and Icy. |
| CytoMAP | https://gitlab.com/gernerlab/cytomap | Software for spatial analysis of segmented cell. |
| Cytomine | https://cytomine.com/ | Web platform that allows for collaborative analysis of large biomedical image collections |
| Fiji/ImageJ | https://fiji.sc/ | Software for biological image analysis with many plugins. |
| Icy | https://icy.bioimageanalysis.org/ | Software for biological image analysis. |
| ilastik | https://www.ilastik.org/ | Interactive tool for image segmentation, classification and analysis. |
| MIB | http://mib.helsinki.fi/ | Software for multi-dimensional image processing, segmentation and visualization. |
| Napari | https://napari.org/stable/index.html | Interactive image viewer for multi-dimensional image in Python. |
| Orbit | https://www.orbit.bio/ | Whole slide image analysis software for digital pathology. |
| QuPath | https://qupath.github.io/ | Whole slide image analysis software for digital pathology. |
| Scikit-image | https://scikit-image.org/ | Python module for image processing. |
| StarDist | https://github.com/stardist/stardist | Deep learning model for image segmentation as a Python module and ImageJ/Fiji plugin. |

### 2.3.2 Cell segmentation using deep learning

Cellular object detection and segmentation is a critical yet challenging step of the conventional feature-based image analysis pipeline. Whereas classical segmentation algorithms such as thresholding and watershed have been commonly used in bioimage analysis software[63], recent advances of deep learning in computer vision have generated various image segmentation models with substantially improved performance[64]. In the 2018 Data Science Bowl, a global competition focusing on 2D nucleus segmentation from high-content images, deep learning approaches dominated the leaderboard, achieving state-of-the-art performance[63]. Many of these deep learning approaches have been incorporated into the open-source bioimage segmentation software[65]. In this section, we first summarize typical deep learning methods applied for cellular object segmentation, and then describe some domain-tailored solutions specifically designed to address the real-world challenges encountered in this field.

*Encoder-Decoder based approaches*

Fully Convolutional Network (FCN) model is among the first deep learning models for semantic image segmentation, comprised solely of convolutional layers to output a spatial segmentation map[66]. Inspired by FCNs, U-Net was developed for segmentation of neuronal structures in electron microscopic stacks[67]. Its architecture comprises two parts: a contracting path to capture context and a symmetric expansive path for precise localization. The contracting path takes an FCN-like structure and extracts features with convolution and down-sampling, while the symmetric expanding path performs up-sampling with deconvolution. To enable precise localization, the feature map from each step of the expansive path is concatenated with the corresponding cropped feature map from the contracting path. A 1x1 convolution operation is used at the final layer to generate the segmentation map[67]. Following the success of U-Net, U-Net variants such as UNet++, a nested U-Net architecture, and UNet3+ have become popular options for medical image segmentation and microscopic image nuclei segmentation tasks[68,69]. In the 2018 Data Science Bowl challenge of nucleus segmentation in 2D microscopic images, the best-performing solution was an ensemble of eight FCNs, six of which used U-Net-like feature decoders[63].

*Multi-Scale and feature pyramid network based approaches*

The concept of feature pyramids is centered around producing multiscale feature representations which integrate semantics at different resolutions. This allows models to detect objects across varying scales. The feature pyramid network (FPN), designed following this principle, serves as a powerful multiscale feature extractor functioning in a fully convolutional manner[70]. Its architecture consists of three primary components: 1) a bottom-up pathway responsible for computing a feature hierarchy consisting of feature maps at multiple downsampled image scales, 2) a top-down pathway designed to generate the feature maps at inversely proportionate, upsampled scales, and 3) lateral connections that fuse two feature maps at identical scaled levels[70]. FPN has been generalized from object detection to instance segmentation task[70]. For example, in the nucleus segmentation competition at the 2018 Data Science Bowl, the second best-performing solution used an FPN-based single neural network model, where the FPN backbone was pretrained on ImageNet—a image database organized by concepts in the WordNet hierarchy[71], and COCO—a large-scale object detection and segmentation benchmark dataset[72]. This solution achieved a good balance of prediction accuracy and inference speed[63].

*Regional convolutional network based approaches*

The Regional convolutional network (R-CNN) approach scans over selective Regions of Interest (RoIs) and use CNNs to classify these RoIs into predefined object categories or a background class[73]. Its extensions such as Faster R-CNN and Mask-RCNN have achieved decent performance in object detection tasks[74,75]. In particular, Mask-RCNN is able to detect objects and generate segmentation mask for each instance simultaneously[75]. When adapted to address the 2D nucleus segmentation challenge, Mask-RCNN pretrained on COCO, has demonstrated competitive performance among the top-ranked solutions[63]. Inspired by this success, an Ensemble Mask R-CNN (EMR-CNN) method has been developed for 3D nuclei segmentation[76]. To bring the R-CNN based approaches to bioimage analysis practice, *Keras R-CNN*, a Python package based on Faster R-CNN, has been implemented in Keras and Tensorflow for cell identification in brightfield and fluorescence images[77]. In addition, CellSeg, a Mask-RCNN based nucleus segmentation software, has been developed for multiplexed fluorescence images[78].

In summary, in the early days of applying deep learning-based models to biological image segmentation tasks, U-Net, FPN and Mask-RCNN models and their variants emerged as leading solutions with decent segmentation accuracy. We refer the interested readers to the report of the 2018 Data Science Bowl results for details in method and performance[63]. Each of these approaches showcases its strengths and drawbacks. For instance, the U-Net model uses skip connections to incorporate feature maps of the whole input image from encoder to decoder, facilitating preservation of global location information and allowing for accurate reconstruction by the decoder[67]. Similarly, FPN also leverages lateral connections, but adopts 1x1 convolution for flexible feature integration, instead of copying and concatenating feature maps from encoder to decoder[70]. In contrast to FCNs that use the full context of the input image, Mask-RCNN works on selected RoIs of an input image to produce the bounding box, class prediction, and segmentation mask simultaneously. This method performs well on instance segmentation tasks to handle multiple objects with complex shapes, albeit more training examples are generally needed compared to U-Net[75].

*Unique challenges for cellular image segmentation and existing solutions*

A common limitation of the aforementioned approaches is that their performances suffer when nuclei are packed densely. In U-Net based models, closely neighbored nuclei are prone to be merged by mistake. In Mask R-CNN based models, nuclei are represented by their axis-aligned bounding boxes, and non-maximum suppression (NMS) technique is used to prevent the algorithm from detecting the same object repetitively. However, when nuclei are crowed, the considerable overlap of bounding boxes can be problematic. To address this challenge, STARDIST was developed to predict a flexible shape representation – a star-convex polygon instead of an axis-aligned bounding box is predicted for each pixel. When benchmarked on the 2018 Data Science Bowl dataset, STARDIST outperformed U-Net or Mask R-CNN based models for intersection over union (IoU) threshold $\tau<0.75$[79]. This method has also been successfully extended for 3D nuclei segmentation (STARDIST-3D)[80]. Fully Convolutional Regression Networks (FCRNs) represents another solution to this challenge, regressing a cell spatial density map of the image. FCRNs demonstrated superior performance at microscopic cell counting when traditional single-cell segmentation fails due to cell clumping or overlap[81]. Another object shape representation approach is proposed by the Cellpose segmentation model. This approach generates topological maps through simulated diffusion and uses human-annotated masks as ground truth. The horizontal and vertical gradients of the topological maps are then predicted to form vector fields. Through gradient tracking, pixels that converge to the same center point are assigned to the same mask[82]. With this representation approach, the Cellpose model outperformed STARDIST, Mask R-CNN and U-Net models at all IoU thresholds on the Cell Image Library dataset[82]. Furthermore, the release of Cellpose 2.0 incorporates an ensemble of diverse pretrained models and human-in-the-loop pipeline to address users' specific needs in cell segmentation[83].

Another limitation of the above-mentioned segmentation approaches is that their training process is fully supervised, thus requiring considerable amount of expert annotations. To alleviate this requirement, Hollandi *et al*. proposed nucleAlzer, which uses image style transfer approaches to generate a set of representative image-label pairs. Applying this data augmentation paradigm to Mask R-CNN based model improved segmentation performance on several image datasets[84].

High computational cost and model complexity were observed when ensemble strategy and sophisticated data augmentation were applied[63]. To mitigate this issue, Dawoud *et al*. introduced a few-shot meta-learning strategy for microscopy image cell segmentation[85]. In this meta-learning algorithm, they combined three loss functions of 1) the standard binary cross entropy loss for pixel-level segmentation, 2) the entropy regularization to shift the segmentation result away from the classification boundary, and 3) the knowledge distillation loss to extract an invariant representation between tasks. Compared to transfer-learning, applying this meta-learning algorithm to U-Net or FCRN architecture significantly improved the mean IoU on fluorescent nuclei images from the BBBC dataset. Further, meta-learning exhibited faster convergence times for both FCRN (40x) and U-Net (60x)[85].

Finally, it is worth noting that the aforementioned models predominantly reply on convolutional neural network. Recently, a novel deep learning architecture, CellViT, was proposed for nuclei segmentation in digitized tissue samples based on Vision Transformer (ViT)[86]. In contrast to CNN-based models, ViTs allow input images with arbitrary sizes and can capture long-range dependencies given the self-attention mechanism[87]. CellViT uses a U-Net-shaped encoder-decoder network, which leverages pre-trained ViTs such as $ViT_{256}$[88] and Segment Anything Model[89] (SAM) as the encoder network and bridges the encoder and decoder components at multiple network depths via skip connections[86]. Although it demonstrated SOTA performance on a histological image dataset[86], it remains to be investigated whether this model can be generalized to the single-cell segmentation task for Cell Painting datasets.

## 2.4 Representation learning for morphological profiling

Feature representation is a critical step in morphological profiling, where the representation can be used for specific downstream tasks, as shown in Figure 4. Morphological features can either be extracted through feature-engineering approach or learned with deep neural network[90]. The former approach can be implemented by a number of image analysis software (section 2.3.1) and has been the classical strategy in morphological profiling pipeline[41]. However, it requires manual efforts in fine-tuning software parameters per experiment setup and relies on expert knowledge to decide what phenotypic features should be measured given the biological system and perturbagen type. In contrast, deep neural networks take an unbiased approach to learn features directly from raw pixels of images and encode meaningful representations[91]. These end-to-end trained deep neural networks not only obviate the need for any segmentation steps, but also the learned representation enables superior performance[28]. Moreover, these networks exhibit improved transferability across different perturbation types (chemical vs. genetic) and demonstrate faster pipeline processing speeds in classification tasks compared to models trained on engineered features[28,92,93].

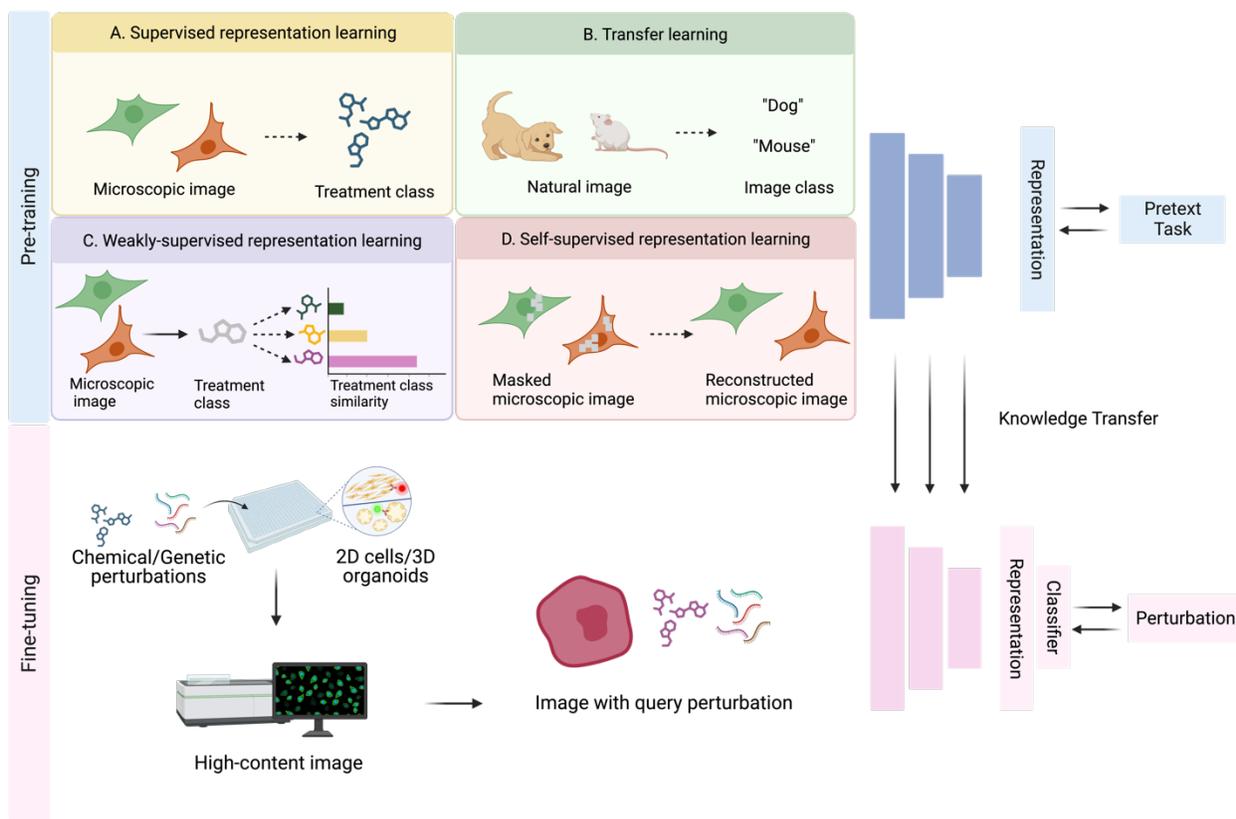

**Figure 4. Representation learning strategies for cell morphology.**
At the pre-training stage, several learning strategies can be applied. **A)** Supervised representation learning employs a deep neural network trained on microscopic image data with the label (treatment class). **B)** Transfer learning utilizes a deep neural network initially trained on other types of annotated image data, such as natural images, to learn representations applicable to microscopic images. The pretext task is to predict the image class. **C)** Weakly-supervised representation learning considers the treatment labels as the weak/noisy labels. A deep neural network is trained on a pretext task to predict the treatment class of the microscopic images. The learned feature embeddings will be used to infer treatment class similarity. **D)** Self-supervised representation learning utilizes the data intrinsic information for model pretraining, such as microscopic image reconstruction. These pretext tasks enhance the model's ability to learn effective representations for major tasks. Following the pretraining stage, the fine-tuning stage transfers the learned knowledge to specific downstream tasks, such as classifying query perturbations to reference perturbations for MOA inference.

*Supervised representation learning*

When extensive annotated training data is available, the utilization of supervised representation learning become particularly effective[94,95]. For example, Kraus *et al.* trained CNNs combined with multiple instances learning on annotated image dataset BBBC021 and yielded higher accuracy in treatment classification compared to the conventional feature-engineering approach[33,94,96,97]. Notably, while this model focuses on the classification task at the whole-image level, it is also capable to perform cell segmentation simultaneously[94]. Similarly, Godinez *et al.*

built a multi-scale convolutional neural network (M-CNN) based classifier, which was trained on the same annotated images[95]. This model outperformed other CNN models on classification tasks when benchmarked on several BBBC datasets.

*Transfer learning*

However, the availability of relevant annotated image data may not always be assured, and the collection of sufficient training data can be expensive and time-consuming. To that end, transfer learning of pre-trained deep neural networks becomes an alternative solution[98]. Pawlowski *et al.* for the first time proposed using ImageNet pretrained CNNs for morphological profiling feature representation and this method achieved superior accuracy and processing speed compared to the feature-engineering based approach[99]. Similarly, Ando *et al.* proposed Deep Metric Network, a model pre-trained on approximately 100 million RGB consumer images, to generate embeddings for the BBBC021 image set[47]. To improve the generated embeddings for representing cellular phenotypes, a feature transformation known as Typical Variation Normalization (TVN) was applied for nuisance variance correction. The yielded Deep Metric Embeddings can robustly represent subtle phenotypic changes by different dosing of perturbagens without any additional training[47]. Many other CNNs pre-trained on ImageNet have also been used to generate cell morphology embeddings[100,101].

*Weakly supervised representation learning*

In addition to transfer learning, weakly supervised learning (WSL) approach has been proposed to train deep neural networks for learning representations of cell painting images[53,54,102]. In this learning schema, treatment or compound labels are treated as 'weak' or 'noisy' labels for several reasons: a) cells may exhibit heterogeneous responses even to the same treatments; b) some treatments are biologically inert; however, in the context of the supervised learning setting with treatments as labels, a deep neural network is nonetheless compelled to identify differences; c) different cell morphology may result from technical artifacts. Therefore, it remains uncertain whether treatment labels accurately reflect cell morphology. To leverage the weak labels, an auxiliary (or pretext) training task is introduced to train a network to classify single cell images to their corresponding treatment labels (the weak labels). Feature embeddings learned from the auxiliary classification task will subsequently be used for the major task, which is to infer the high-level associations between treatments based on similarity. In the setting of drug discovery, this allows for MOA prediction through assigning query compounds to a library of annotated reference compounds[53,54,102,103]. Given that these deep neural networks are exposed to the distributions of both true biological phenotypes and confounding factors in the pretext training task, disentangling phenotypes from confounding factors is crucial to the success of this training schema. To achieve this, besides batch correction efforts (summarized in section 4), a few other strategies have proven to be helpful, such as RNN-based regularization[102], convex combinations of images to generate new samples[102], and combining image datasets with strong perturbations for training[53]. Beyond representation learning with broadly used CNNs, WS-DINO from Cross-Zamirski *et al.* was proposed to learn representations using a knowledge distillation approach with ViT backbone. In this approach, global and local crops from different images under the same treatment are generated[54]. The teacher network is exposed solely to global crops, whereas the student network sees both, and the objective is to minimize the cross-entropy loss between student and teacher prediction output. Notably, in contrast to many other WSL approaches, WS-DINO does not require single cell cropping for pre-processing[54].

*Unsupervised representation learning*

Finally, unsupervised learning approaches provide another avenue for feature representation learning by identifying underlying patterns in raw data or clustering similar data into groups. Examples of such exploitable unlabeled information include whether images belong to the same treatment[104], metadata information[105], and pseudo-labels assigned by K-means clustering on embeddings[106]. Another strategy is to use generative models[107] such as GANs[49] or VAE framework[50,51] to learn feature representations. They function by learning and generating new data distributions that are similar to the training data, thereby learning inherent structures and patterns within the dataset. In addition, self-supervised learning (SSL) approach can use a pretext training task, mining the intrinsic information present in the data itself, to train a CNN capable of learning effective feature representation and use it for downstream analysis[108]. For the pretext task, Lu *et al.* proposed 'paired cell inpainting', whereby the model needs to identify protein localization from the 'source' cell and predict the similar localization in the 'target' cell[108]. Contrastive loss-based approach can also learn robust cell representations by training the model to bring positive example representations closer in the feature space and push the negative example representations further away from the positive ones[109]. Perakis *et al.* demonstrated that representations learned with the contrastive learning framework can be used in MOA classification task with the impressive performance on par with the transfer learning approach[47,109]. Beyond the CNN-based model, the SSL method has also been employed to pre-train the ViT architecture, resulting in significant enhancements even in segmentation-free morphological profiling[28,110]. In evaluations using subsets of the JUMP-CP consortium data, the ViT architecture, trained by the recently introduced DINO SSL approach, outperformed both CellProfiler and transfer learning-based methods in several dimensions. Specifically, when trained on multisource data (datasets generated by multiple institutions), this approach demonstrated the best performance in classification tasks. The resultant image representations showcased exceptional adaptability, transitioning efficaciously from chemical to genetic perturbations. Moreover, the pipeline functioned at speed 50 times faster than CellProfiler-based feature engineering workflow[28]. It is noteworthy that, unlike CNNs where local features are consolidated into aggregated vectors, ViTs preserve a more refined resolution of inputs across all network layers, and this preservation facilitates the encoding of features that are biologically meaningful at the subcellular level[110]. Notably, ChannelViT has been proposed to make a simple modification to the ViT architecture by constructing patch tokens independently from each input channel and includes a learnable channel embedding. These modifications improve model reasoning across channels, such that the model can generalize efficiently even when limited input fluorescent channels are available. When trained with DINO algorithm, ChannelViT consistently outperforms standard ViT on input images with varying sets of fluorescent channels[111]. Altogether, these findings underscored the formidable efficacy and robustness of SSL approaches in morphological profiling.

Most of the representation learning approaches described in this section have been benchmarked on the BBBC021 dataset with these evaluation metrics. Their performance is summarized in Table 3. From this comparison, WS-DINO, the weakly-supervised method from Cross-Zamirski *et al.*[54] achieved the best performance. The transfer learning method from Ando *et al.*[47] and the self-supervised contrastive learning method from Perakis *et al.*[109] also showcased strong performance in learning meaningful phenotypic embeddings.

**Table 3. Model performance comparison by MOA classification accuracy on the BBBC021 dataset.**

| Approach | Description | NSC* | NSCB† | Drop‡ | Reference |
|---|---|---|---|---|---|
| Conventional feature engineering | CellProfiler with Factor Analysis | 94% | 77% | 17% | Ljosa 2013[96] |
| | CellProfiler with illumination correction | 90% | 85% | 5% | Singh 2014[97] |
| Supervised learning | CNN with Noisy-AND pooling function | 96% | N/A | N/A | Kraus 2016[94] |
| | Multiscale-CNN | 93% | N/A | N/A | Godinez 2017[95] |
| Transfer Learning | ImageNet Pretrained Inception-v3 with illumination correction and greyscale transformation | 91% | N/A | N/A | Pawlowski 2016[99] |
| | Pretrained Deep Metric Network with TVN postprocessing | 96% | 95% | 1% | Ando 2017[47] |
| Weakly supervised learning | Weakly supervised ResNet-18 with Mixup regularization | 95% | 89% | 6% | Caicedo 2018[102] |
| | WS-DINO finetuned on BBBC021 with compound as weak label | **98%** | **96%** | 2% | Cross-Zamirski 2022[54] |
| Self-supervised learning | CytoGAN (LSGAN) | 68% | N/A | N/A | Goldsborough 2017[49] |
| | VAE+ | 93% | 82% | 11% | Lafarge 2019[50] |
| | UMM discovery with NSCB as best epoch criterion | 95% | 89% | 6% | Janssens 2021[106] |
| | Contrastive learning with whitening postprocessing | 96% | 95% | 1% | Perakis 2021[109] |

\* **NSC (Not-Same-Compound matching accuracy):** In the NSC setting, one compound is deliberately excluded during the training phase and the model is subsequently tasked with the excluded compound's treatment class, given all other treatments. This is achieved by using a 1-Nearest-Neighbor (1-NN) classifier to assign the excluded compound to the nearest neighbor in the feature space, and none of the neighbors (known compounds) belong to the same

compound treatment. This metric is to evaluate the model's capacity to adequately generalize and correctly infer a new compound's treatment class when its MOA is unknown[96].

† **NSCB (Not-Same-Compound-and-Batch matching accuracy):** NSCB serves as a more stringent metric compared to NSC. By excluding a compound and a full experiment batch during the training process, this metric enables a more robust evaluation of model's performance and generalizability across different experiment conditions and batch settings. This metric reflects the impact of batch effects and other confounding factors[47].

‡ **Drop:** Drop is calculated by subtracting NSCB from NSC. Ideally, performance drop should not be observed. The larger this metric value is, the more substantial the batch effect is[102].

## 2.5 Integrating morphological data in multimodal learning for drug discovery

With the advances in biotechnology, a wealth of data from various modalities can be generated and collected to facilitate drug discovery. Cheminformatics, for example, has made substantial contribution to drug discovery through analysis and representation of chemical structures and exploiting the similarity principle[112]. Chemical structure data of compounds is always readily available, and predicting compound bioactivity based on this data modality can be performed virtually. However, elucidating the intricate relationship between structure and biofunction is a challenging task[112]. On the other hand, 'Omics' profiles, such as genomics, transcriptomics, proteomics and metabolomics, can characterize treatment outcomes from different aspects. However, assay cost and scalability emerge as major concerns for high-throughput studies[113]. Indeed, every modality of data utilized in the drug discovery presents its unique set of advantages and disadvantages. A detailed comparison is summarized in Table 4. Integrating these modalities is promising to maximize their potentials and mitigate the limitations, thereby providing a comprehensive understanding of treatment effects. Notably, recent research has shown that different data modalities, such as chemical structure, morphology, and gene expression, exhibit complementary strengths in predicting treatment effects[114]. Integrating morphological data with other data modalities using machine-learning or deep-learning based approaches has now become an active field of research.

**Table 4. Comparison of transcriptomic and morphological profiling data for drug discovery.**

| Attribute | Morphological profiling | Transcriptomic profiling |
| --- | --- | --- |
| Infrastructure requirements | High-content imaging system. Some requires lab automation workflow. | Next-generation sequencer. Some requires cell-sorting capability. |
| Scalability | Scalable for cell painting assay. | Scalable for L1000 assay. |
| Cost | Low cost for conducting assays but high cost in system setup. | In general, low cost for newer platforms. |
| Data interpretability | Not interpretable on gene expression level. | Interpretable on gene expression level. |
| Data processing framework | Best practices for conventional feature-engineering approach have been made. Processes such as batch correction remain to be standardized. | Mostly standardized. |
| Reproducibility | Can be experimental platform dependent. Variations between data producing sites is non-trivial. | Technically reproducible. Biological reproducibility usually needs to be confirmed. |

The integration of structural models with cell morphology models has been demonstrated to improve biological assay outcome prediction accuracy. Seal *et al.* proposed the similarity-based merger model, which combines the scaled predicted probabilities from individual models trained on Cell Painting images and chemical structures, and the morphological and structural similarities between test and training compounds[115]. Specifically, the predictions from individual models and similarity values are used to fit a logistic regression model to predict the test compound activity. The authors demonstrated that the similarity-based merger model outperforms soft-voting ensemble, hierarchical model, or either of the individual models trained on unimodal data[115].

In addition, SSL techniques such as contrastive learning approaches have also been utilized to align multimodal data sources to enrich morphological profiling analysis in drug discovery[116-118]. For example, a method known as Contrastive Leave One Out boost for Molecule Encoders (CLOOME) has been proposed, aiming to learn aligned representations derived from the compound's chemical structure and the corresponding cellular images obtained after treatment with the same compound[116]. Its learning framework incorporates a microscopy image encoder, a molecule structure encoder, and uses the InfoLOOB objective[119] to learn the aligned embedding of treatment image and compound structure[116]. Similarly, Zheng *et al.* presented Molecular graph and hIgh content imaGe Alignment (MIGA) framework with an image encoder and a graph neural network (GNN) based structural encoder[118]. To align graph embeddings with image embeddings, three contrastive objectives are used: graph-image contrastive learning, masked graph modeling and generative graph-image matching. The crossmodal representation learned with this framework improves performances on several downstream tasks[118]. This approach is extended further by Nguyen *et al.* to develop Molecule-Morphology Contrastive

Pretraining (MoCoP)[117]. This framework uses a morphology encoder, a gated graph neural network (GGNN) based molecule encoder and the modified InfoNCE objective[120] to learn multimodal representation. The GGNN pretrained with MoCoP can be finetuned for downstream quantitative structure-activity relationship (QSAR) tasks[117]. Furthermore, active learning approach has been used to boost the performance of image-based and structure-based models and benefit the downstream QSAR tasks. The initial image-based and structure-based models assist selecting candidate compounds to be validated in toxicity assays. Once the wet-lab assays are completed, assay readouts will be collected as new annotations to continue refining both models. This iterative approach has been applied to detect compounds with mitochondrial toxicity[121].

In addition to chemical structure data, integrating transcriptomic profile with cell morphology serves as another crossmodal combination. A prevalent assay for obtaining gene expression profile is the L1000 assay[122]. Both Cell Painting and L1000 assays are scalable and provide complementary data. Compared to the transcriptomic profile from L1000, the morphological profile from cell painting is more reproducible yet susceptible to batch and well position effects. Conversely, L1000 captures more diverse features. Collectively, these two profiling modalities measure overlapping and assay-specific MOAs[123]. Besides L1000 transcriptomic profile, another gene expression-based assay, Functional Signature Ontology (FUSION), can be fused with morphological profiling data to assign MOAs to complex natural product fractions in pair with metabolomic profiling data[124]. Comparative studies have shown that transcriptome-based and morphology-based models offer comparable or better performance in MOA prediction, compared to chemical structure-based model[125]. These findings provide rationale and potential advantages of integrating transcriptomic and morphological profiling for drug discovery. More discussions on the applications and concerns of integrating these two data modalities have been recently characterized[126,127]. Datasets with matched transcriptomic and morphological profiling data have been summarized in Table 5.

**Table 5. Multimodal datasets with matched transcriptomic and morphological profiling data.**

| Cell type | Transcriptomic profiling description | Transcriptomic profile URL/Identifier | Morphological profile URL/Identifier | Reference |
|---|---|---|---|---|
| A549 | L1000 | https://figshare.com/articles/dataset/L1000_data_for_profiling_comparison/13181966/2 | https://idr.github.io/idr0125-way-cellpainting/ | Way 2022[123] |

| A549 | L1000 | https://www.ncbi.nlm.nih.gov/geo/query/acc.cgi?acc=GSE83744 | https://registry.opendata.aws/cell-painting-image-collection/ | Haghighi 2022[127] |
| --- | --- | --- | --- | --- |
| A549 | L1000 | https://figshare.com/articles/dataset/L1000_data_for_profiling_comparison/13181966 | https://zenodo.org/records/3928744#.YNu3WzZKheV | Haghighi 2022[127] |
| U2OS | L1000 | https://www.ncbi.nlm.nih.gov/geo/query/acc.cgi?acc=GSE92742 | https://idr.openmicroscopy.org/webclient/?show=screen-1251 | Haghighi 2022[127] |
| U2OS | L1000 | https://www.ncbi.nlm.nih.gov/geo/query/acc.cgi?acc=GSE92742 | http://www.cellimagelibrary.org/pages/project_20269 | Haghighi 2022[127] |
| U2OS | L1000 | https://github.com/carpenterlab/2017_rohban_elife/tree/master/input/TA-OE-L1000-B1 | https://idr.openmicroscopy.org/webclient/?show=screen-1751 | Haghighi 2022[127] |
| Hela | FUSION | Upon request | Upon request | Hight 2022[124] |

Data fusion methods have been widely used to integrate multimodal data (Figure 5). In general, these methods can be categorized as early fusion and late fusion. Early fusion works by integrating the separate raw data modalities into a unified representation before feeding into the deep learning model for feature extraction. In contrast, late fusion combines the predictions of individual models, each built on a specific data modality. Algorithms such as cooperative learning have been proposed to enhance the alignment between predictions[128]. To integrate morphological, transcriptomic and chemical structure profiles, Seal *et al.* compared both early and late fusion methods in detecting mitochondrial toxicity. They reported that late fusion model can accurately determine the mitochondrial toxicity of compounds that have inconclusive toxicity results reported previously[129].

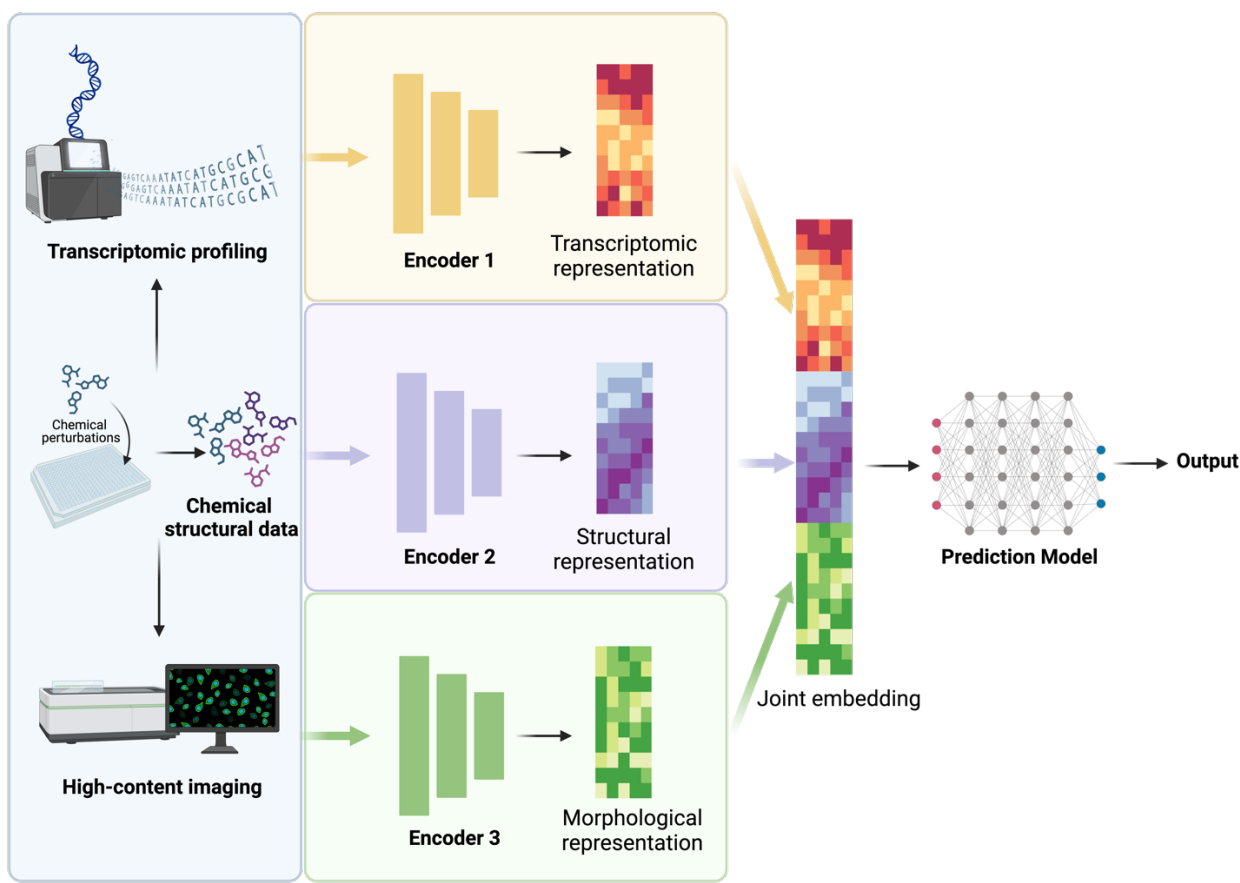

**Figure 5. Combine morphological data with other data modalities.**
The image data obtained from morphological profiling assays can be combined with other modalities of profiling data to perform downstream tasks jointly. One strategy involves training individual models to extract representations from each data modality, such as image data, chemical structural data, and transcriptomic data. These individual representations contribute to a joint embedding, which is subsequently utilized for downstream analyses.

To identify perturbation effects in distinct feature space of morphological and transcriptomic data, Smith *et al*. proposed Perturbational Metric Learning (PeML) for similarity metric learning for multimodal data representation[130]. This WSL approach aims to learn an embedding to maximize the similarity between replicates while non-replicates stay dissimilar. This learning methodology can be applied to both morphological and transcriptomic profiles and has demonstrated improved performance in MOA prediction[130].

In summary, applying deep learning approaches to integrate morphological data with other modalities, such as chemical structure and transcriptional data, demonstrates growing importance in drug discovery efforts. Techniques like contrastive learning and various data fusion methods are emerging to align multimodal data. The continuous curating of such multimodal datasets will further contribute to this burgeoning field.

## 3 Novel applications of morphological profiling in drug discovery

Machine and deep learning approaches have significantly contributed to morphological profiling, enriching various aspects of phenotypic drug discovery. Applications such as identifying small molecule MOAs, lead optimization, and predicting toxicology have been extensively reviewed elsewhere[131-134]. In the following sections, we will discuss the recent advances in several novel applications.

### 3.1 Construct genotype-phenotype relationship and gene function network.

Mapping genotype to disease-relevant phenotype has been a critical question in genomics. To address this challenge, genome-scale pooled CRISPR screens have been used to provide insights into gene functions. However, conventional screening readouts are relatively low in dimensionality (such as cell viability, proliferation, or expressions of biomarkers), thereby providing a constrained view of disease-relevant phenotype[135]. While high content transcriptomic data from scRNA-seq can be meansured from pooled CRISPR screens, the cost of achieving high content readout as such from a genome-wide CRISPR screen can be unfeasibly high[135]. To overcome this hurdle, image-based profiling can provide high content morphological readout for CRISPR screens at genome-scale[38,136,137]. Notably, optical pooled CRISPR screens[138] can be combined with image-based profiling to create a genome-wide perturbation atlas and to construct a gene function network based on the uncovered genotype-phenotype relationships[136,139,140]. For example, Ramezani *et al.* developed a cell painting based optical pooled cell profiling approach (PERISCOPE) to allow pooled CRISPR screens to have high-dimensional cellular morphological profiles as endpoint readouts. This scalable pipeline has been applied to A549 cells and human cervical cancer cells (HeLa) to investigate gene knockout responses and identify gene clusters based on morphological similarity[139]. Sivanandan *et al.* introduced a similar technique termed Cell Painting Pooled Optical Screening in Human Cells (CellPaint-POSH). With this approach, a screening with a druggable genome library of 1640 genes has been conducted on A549 cells. Notably, this work applied the SSL DINO-ViT model (section 2.4) for image representation and demonstrated decent performance in recovering gene function network[140]. Such results further attest to the efficacy and robustness of deep learning approaches in generating informative image representations, subsequently leading to valuable biological insights.

In the efforts of mapping genotype to phenotype, the observation of "proximity bias" has been reported, whereby the phenotypes of CRISPR knockouts demonstrate higher similarity to biologically unrelated genomically proximal genes on the same chromosome arm than the biologically related genes. The cause of this artifact arises from widespread chromosome arm truncation due to Cas9 nuclease activity and is not observed in shRNA or CRISPR interference (CRISPRi) perturbations. Performing arm-based geometric normalization of features at gene level can reduce this bias without compromising the recovery of biological relationship[141].

## 3.2 Characterize perturbation impacts in dynamics.

An emerging advance of morphological profiling is towards live-cell phenotyping, which can be performed by fluorescent or phase-contrast imaging, and by continuous imaging[32,142] or dynamic imaging[51]. Several advantages accompany this approach. First, adding temporal variables to morphological profile improves assay predictive power[32]. For example, in a live-cell imaging-based profiling assay, a library of 1,008 FDA approved drugs with manual annotations was profiled against 15 reporter cell lines that expressed fluorescent protein-tagged organelle or pathway markers. The morphological profile was generated from 24-hour high-content imaging and can be used to accurately infer 41 of 83 testable MOAs[32]. Beyond this, live-cell imaging enables the characterization of cell state transition dynamics, a critical feature in developmental biology[51,142]. Human pluripotent stem cells (hPSCs) co-expressing histone H2B and cell cycle reporters can be profiled in a multi-day, high-content manner at single-cell resolution. With this profile, a deep learning model can be trained to provide highly sensitive predictions of spatiotemporal single-cell fate dynamics, as early or even earlier than cell state-specific reporters[142]. Moreover, live-cell morphological features of human induced pluripotent stem cells (hiPSCs) can even be used to predict differentiation marker gene expression[51]. This approach involves performing phase-contrast imaging and bulk RNA-sequencing at each consecutive passage of hiPSCs. A VAE variant, VQ-VAE[143], learns the image feature vector in a self-supervised approach. A number of Support Vector Regression (SVR) models, each corresponding to a differentiation marker, were trained to predict differentiation marker gene expression from the image feature vector. Bulk RNA-sequencing readouts were used as labels for this supervised learning process. Altogether, this approach builds the relationship between transcriptional and live-cell morphological profiles[51].

Deep learning models such as DynaMorph[144] and DEEP-MAP[142] have been proposed to analyze morphological profiles in dynamics. To take DynaMorph for example, VQ-VAE was trained to learn a representation of cell shape through a self-supervised image reconstruction auxiliary task. To ensure that cell shape changes smoothly between neighboring frames, a temporal matching loss was applied. The representation of cell shape regularized by the temporal continuity can distinguish morphodynamic states of microglia in response to pro- and anti-inflammatory stimuli[144].

### 3.3 Guide *de novo* hit design.

Although the typical downstream applications of morphological profiling have been focused on clustering or classification tasks (section 2.2), Zapata *et al*. proposed to leverage morphological profiles to guide *de novo* molecular design with GANs[145]. Compared to using transcriptional profiling for compound *de novo* design[146], morphological profiling provides higher throughput with less cost. More importantly, more than 40% of the generated molecules have drug-like physicochemical properties, and more than half are expected to be synthesizable. This model can also be generalized to morphological profile with genetic perturbations such as gene overexpression. These findings indicate that this approach is able to effectively translate morphological similarity into chemical similarity with high efficiency[145].

### 3.4 Facilitate image-based profiling in advanced biological models.

Organoids are hetero-cellular biomimetic tissue models that has become a powerful experimental tool transforming basic science and translational research[147]. While the traditional low throughput methods provide valuable biological insights high throughput methods are needed to fully exploit the potential of organoids as ex vivo models. Modelling the development of disease with organoids that can recapitulate tissue structure, pathology, phenotypes and differentiation has revolutionized the study of various human disease including cancer[147,148]. In a recent study we demonstrated the effect of small molecules in mouse pancreatic acinar that causes inhibition or reversal of acinar-to-acinar ductal metaplasia (ADM) using high-content image-based screening in organoid culture[149,150]. Advances in technology in organoid culture and the remarkable self-organizing properties reflecting key structural and functional attributes of organs such as brain, kidney, lung, gut or similar even hold promise to predict drug response in a personalized fashion.

While organoids are normally cultured in bulk in an extra cellular matrix, these bulk cultures can physically overlap which makes it challenging to track the growth and properties of individual organoids in high-throughput assays. Various microwell designs have been introduced to overcome specific challenges associated with image-based analysis but still struggle with large numbers of organoids[151,152]. Using different organoid culture methods, phenotypic assays can be designed using features like whole organoid morphology, growth rates or movement with simple brightfield imaging. Many of these methods rely on cellular aggregation to generate spheroids rather than growing organoids from single cells[153-155]. These can cause limitations in understanding the phenotypic heterogeneity while most of the methods do not employ integrated analytical pipelines into overall workflow[156-160] or the ability to selectively retrieve organoids for downstream investigations. Overcoming some of these issues Forsyth and a team of researchers[147] has built an open source microwell-based platform for high-throughput quantification using image-based parameters. The method utilizes an organoid-optimized deep-learning model that can be integrated with existing culturing protocols and micro-well platforms to investigate phenotypic features across different tissues. Additionally, patient derived tumor organoids have been developed into powerful organoid based discovery platforms in recently demonstrated using CRISPR-Cas9 screening for patient-specific functional genomics[161]. Defined mutations are introduced to transform normal organoids to tumorigenic growth upon xenotransplantation, combining the exploratory power of CRISPR-Cas9 screening with 3D organoids[154,157,162]. These advances demonstrate that organoids are powerful experimental models for morphological profiling to study the maturation and progression of various diseases.

**3.5 Enable natural product-based drug discovery.**

Natural products (NPs) and their structural analogues have made a major contribution to pharmacotherapy, playing a key role in drug discovery[3,4]. Recent years have witnessed AI approaches have substantially advanced the efficient identification of drug candidates from NPs, marking notable progress in drug discovery[5]. NP-based drug leads are typically identified by phenotypic assays[4]. To that end, imaged-based profiling platform has been developed to study toxicity, structure-activity relationship (SAR), MOA and potential off-target effects of NPs[6]. For example, a high-throughput screening on MIN6 β cells with 6,298 marine NP fractions has been performed to select for hit compounds with nontoxic and long-lasting effects in inhibiting glucose-stimulated insulin secretion[163]. In combination with mass-spectrometry analyses and NMR analyses, aureolic acid CMA2 has been identified as the major component of the top hit fraction derived from *S.anulatus*. Treating MIN6 cells with CMA2 leads to decreased nuclei counts determine by DAPI staining, attesting to its bioactivity[164]. In another study, botanical NP extracts have been screened for blockade of SARS-CoV-2 infection in human 293TAT cells. A leading hit, the extracts of *S. tetrandra*, is further investigated on its anti-viral MOA through phenotypic assays based on intracellular phospholipids formation[165]. In addition, high-dimensional phenotypic readouts also assist exploring NP MOA. To understand the MOA of the Polyketide Lagriamide B from the Burkholderiales strain, its morphological impact on U2OS cells is investigated through the Cell Painting assay followed by high-content imaging. At low treatment concentration,

Lagriamide B leads to disruption in actin polymerization and incomplete cytokinesis, and at high concentration, low cell count and decreased cell size are observed. These phenotypic effects indicate an MOA of Lagriamide B in actin polymerization disruption[166].

Specifically, integrating morphological with multi-omics profiling helps annotate the bioactive components of NPs, which addresses one of the most significant challenges in NP-based drug discovery[124]. For example, an integrated framework of morphological and transcriptomic profiles has been used to annotate mass spectrometric profile of marine bacteria extracts[124]. This orthogonal platform demonstrated a new paradigm to understand the association between NP components and treatment phenotypes and underscored the importance of integrating multimodal profiling data for drug discovery (section 2.5).

## 4 Challenges and Outlook

Morphological profiling is poised to have a profound and continuing impact on phenotypic drug discovery in the next decade and beyond. As advances in AI permeate various aspects of drug development and discovery[167], machine learning and deep learning approaches will continue to empower morphological profiling with enhanced accuracy and efficiency[131]. However, several challenges await resolution in order to fully leverage cellular images as a reliable and insightful resource, as will be discussed in this section.

Although representation learning (section 2.4) has become a robust approach to learn cellular features with less manual input than conventional feature-engineering approach, it is susceptible to confounding factors such as batch effects. Batch effects are variations in data caused by the differences in the technical execution of each experimental batch. Such confounding factors introduce irrelevant sources of variation into data, and can potentially mislead biological conclusions[40]. Disentangling these confounding factors from phenotypes is a crucial step to recover true biological signal. Significant progress has been made in this regard, with methods such as TVN[47], BEN[168], TEAMS[169], CDCL[105], and GRU-based regularization[102]. Furthermore, batch correction methods for transcriptomic profiles may be applicable. A recent study on subsets of JUMP-CP demonstrated that Harmony, a non-linear method developed for processing scRNA-seq data, consistently outperforms other transcriptomic profile batch correction strategies in balancing batch removal and biological variation conservation[170]. In addition to the aforementioned methods, adding a context token to include batch-specific information during image representation learning also demonstrated decent performance in out-of-distribution generalization and batch variation handling[171]. To evaluate and compare batch correction strategies, RxRx1, a Cell Painting image dataset of genetic perturbations with 51 experimental batches from four cell types, has been systematically designed[40]. With the development and sharing of the benchmarked dataset, future work will continue to enhance upon existing methods. Improved handling of the confounding factors will further facilitate data sharing and reproducibility between data generation sites, thereby bringing significant benefits to the broader scientific community.

The success of phenotypic drug discovery heavily relies on disease relevance of the biological model. Applying relevant cell types and perturbations in morphological profiling assay is essential, but not sufficient to guarantee translatability[1]. Recent efforts have been made to apply increasingly multiplex biological model systems for image-based profiling, such as co-cultured 2D cell lines[172] and 3D organoids[173,174]. However, on the computational side, most approaches have been built upon Cell Painting assay images from mono-cultured 2D cells. Therefore, many challenges remain in generalizing these approaches to a multiplex biological model. For example, how do current cell segmentation (section 2.3.2) and representation learning methods (section 2.4) perform on 3D images? How is the quantity and quality of 3D image dataset which can be utilized for effectively training for fine-tuning deep learning models? How generalizable are the representation learning frameworks (section 2.4) to cellular images consisting of multiple cell types, each demonstrating different morphology? How to integrate morphological data and other modalities of data (section 2.5) from a multiplexed cell system to obtain cell-cell interaction information? Overcoming these hurdles will bring morphological profiling to the next level of clinical translatability.

In terms of integrating morphological profile with omics data (section 2.5), compared to bulk transcriptomic readouts, single-cell transcriptomics, spatial transcriptomics and translatomics offer a wealth of gene expression information at individual cellular and subcellular levels[175-178]. Advances such as sci-RNA-seq3 have enabled single cell transcriptional profiling in high throughput[179]. Given this technical progress, future work may establish an orthogonal profiling platform to combine morphological and single-cell profiling, thereby linking molecular phenotype to cellular phenotype at single-cell resolution. In addition, integrating Perturb-seq with image-based profiling will become a promising future direction to characterize the impact of genetic perturbations with single-cell transcriptomics and morphological readouts[180]. High quality datasets of such should be established to encourage the development and evaluation of data integration approaches.

Last but not least, despite the impressive inferential capabilities of deep learning approaches, drawbacks remain that the explainability of these 'black box' models is unsatisfying[181]. In drug discovery especially, model interpretability is important to ensure that the biological conclusions are valid. To mitigate this, several efforts have been initiated to improve model interpretation in morphological profiling. For example, Chow *et al.* trained VAEs to interpret latent space feature representations in Cell Painting assay[182]. In the broader field of computer vision, techniques such as class activation mapping[183,184] have been proposed to provide visual explanations for deep neural networks. Future work should continue to develop or advance techniques as such to morphological profiling to enhance model interpretability[183].

## 5 Concluding remarks

Morphological profiling represents a powerful, high-throughput, data-intensive, and cost-efficient technique for phenotypic drug discovery. It offers an unbiased and high-dimensional image readout of cellular phenotype in response to various perturbations, thereby providing a comprehensive view on compound bioactivity. Emerging techniques from computational biology and deep learning communities have made significant progress in enhancing the analytical pipeline from representation to prediction. While challenges remain in this fast-evolving field, future work will continue to coordinate multidisciplinary efforts in leveraging visual phenotypes to empower drug discovery.

## Acknowledgment

Y. Li is supported by the University of Florida (UF Startup Fund). H. Luesch is generally supported by the Debbie and Sylvia DeSantis Chair professorship, and for phenotypic screening, HCS and morphological profiling supported by NIH grants R01CA172310 and RM1GM145426. C. Li has been partially supported by the Bodor Professorship Fund, UF AI Catalyst Fund and NCI R01CA212403. J. Bian is supported by the following NIH grants, R01AG076234, RF1AG077820, and UL1TR001427. Y. Ding is partially supported by NIH R35GM128742. L. Cui is supported by the University of Florida (UF Startup Fund). We thank Wenjun Xie and Yuzhao Zhang for helpful discussion and comments that improved the manuscript. Figures 1,3,4 and 5 were created with BioRender.com.